\documentclass[10pt,aps,twocolumn,groupedaddress,showpacs]{revtex4}
\usepackage{graphicx,amsmath,amsfonts}

\begin{document}
\title{Reflection of light and heavy holes from a linear potential
barrier.}

\author{Anatoli Polkovnikov}
\email{anatoli.polkovnikov@yale.edu}
\homepage{http://pantheon.yale.edu/~asp28}
\altaffiliation{also at Ioffe Physico-Technical Institute}
\affiliation{Physics Department  Yale University}

\author{Robert A. Suris}
\email{suris@theory.ioffe.rssi.ru}
\homepage{http://www.ioffe.rssi.ru/Dep_TM/suris.html}
\affiliation{Ioffe Physico-Technical Institute}

\date{\today}

\begin{abstract}
In this paper we study reflection of holes in direct-band semiconductors
from a linear potential barrier. It is shown that light-heavy hole
transformation matrix depends only on dimensionless
product of the light hole longitudinal momentum and the characteristic
length determined by the slope of the potential and it doesn't depend on the
ratio of light and heavy hole masses, provided this ratio is small.
This coefficient is shown to vanish both in the limit
of small and large longitudinal momenta, however the phase of a reflected
hole is different in these limits. An approximate analytical expression for
the light-heavy hole transformation coefficient is found.
\end{abstract}

\pacs{7155.Eq, 7210.Fk}

\maketitle

\newsymbol\rightleftarrows 131D

\section{Introduction}
Direct band A$_3$B$_5$ semiconductors are very challenging both for
theoretical and experimental applications. In particular,
heterostructures based on these compounds are widely studied. There have
been published many papers devoted to the boundary condition problem in
heterostructures~\cite{Dyak,Sur1,Chuang,For1,Kumar,For2,Kisin}. These
articles mainly dealt with abrupt interface potentials, where
appropriate boundary
conditions were sought to relate the wave functions at the different sides
of the interface. The most common approach used to find the boundary
conditions is to integrate envelope function equations through the
interface~\cite{Sur1,Chuang,For1,For2}. In this way, it appears that there
is no mixing between light and heavy holes at normal incidence~\cite{Chuang}.
However, there is no general reason justifying the integration procedure.
Moreover, the tight binding model and some numerical calculations
(see ~\cite{Kaminski} and references therein)
are in direct contradiction with the above statement. In particular, for
a GaAs/AlAs interface the light-heavy hole transformation coefficient for
the normal incident flux was found to be of order unity~\cite{Kaminski}.
In a purely phenomenological analysis, relying on group theoretical
formalism~\cite{Kisin}, the boundary conditions were found to be determined
by a number of phenomenological parameters which should be found
either from microscopical calculations or from fitting to experimental
data. However, this approach hardly allows following the dynamics of the
reflection matrix as a function of energy and lateral momentum of incident
holes. On the other hand there is possible a completely different
situation wherein the potential near the turning point is smooth.
The boundary of this type may appear
either in a compound with relatively slow variation of composition or
in a structure subjected to an external electric field. Smoothness of the
potential justifies use of the envelope function equations and hence
eliminates boundary condition uncertainties.

The formulation of the problem suggests that the semiclassical approach
is to be used. However, being applied to Luttinger equations, it
gives new essential features as compared to standard WKB formulas. The
situation of a smooth linear potential was first considered in~\cite{Volk}
for a gapless system, and later, in~\cite{Suris}, for the case of A$_3$B$_5$
semiconductors. In the latter paper, it was shown that there is a
light-heavy hole transformation even in a slow varying potential if the
incident particle flux is not normal to the interface. In their calculation,
the authors used an infinitely large heavy hole mass, which resulted
in divergence of the wavefunction in a certain underbarrier point at which
the kinetic energy of a hole was equal to zero. In the present paper we
take the ratio of heavy and light hole masses to be large but finite
and avoid divergences of this type. We note finally
that though we assume that the hole masses and spin-orbital splitting
don't depend on the coordinate, their slow linear variations can
be easily included into the model.
\vspace{0.05cm}
\section{Solution of Luttinger Equations in a Linear External
Potential}
\vspace{0.1cm}
Let us start from the conventional $8\times 8$ Hamiltonian in a spherical
band approximation, i.e. assuming that the Luttinger parameters
$\gamma_2$ and $\gamma_3$ are equal to each other~\cite{Sercel,Polk}. This
assumption is usually well justified in the vicinity of $\Gamma$-point.
Instead of total angular momentum basis, we use another one~\cite{Polk,For2}
\begin{equation}
|s\uparrow\rangle,\;\, |s\downarrow\rangle,\;\,
|x\uparrow\rangle,\;\, |x\downarrow\rangle,\;\,
|y\uparrow\rangle,\;\, |y\downarrow\rangle,\;\,
|z\uparrow\rangle,\;\, |z\downarrow\rangle,
\end{equation}
which allows us to write the equations for electron and hole wave
functions in a compact analytical form:
\newpage
\begin{widetext}
\begin{eqnarray}
&&(E_g+\phi(x)-E)\Psi_s-i\gamma \nabla {\bf \Psi}=0\nonumber\\
&&\left(-E+\phi(x)-{\Delta_{so}\over3}\right)
{\bf \Psi}-i\gamma\nabla\Psi_s+
{1\over 2m}(\tilde{\gamma_1}+4\tilde{\gamma_2})\nabla
(\nabla{\bf \Psi})
-{1\over 2m}(\tilde{\gamma_1}-2\tilde{\gamma_2})
\nabla\times[\nabla\times{\bf \Psi}]+i{\Delta_{so}\over 3}
[{\bf \sigma}\times{\bf \Psi}]=0,
\label{eq:1}
\end{eqnarray}
\vspace{-0.2cm}
\end{widetext}
where $\tilde{\gamma_1}$ and $\tilde{\gamma_2}$ are generalized Luttinger
parameters~\cite{Sercel}, $\gamma$ is the Kane parameter having dimension of
velocity, $m$ is a free electron mass,
${\bf \sigma}=(\sigma_x,\sigma_y,\sigma_z)$ are Pauli matrices,
$E_g$ and $\Delta_{so}$
are the band gap and spin-orbital splitting, $\phi(x)$ is the external
potential which will be considered hereafter to be a linear function of
$x$-coordinate: $\phi(x)=\alpha x$. We use the units where Plank constant
is equal to unity. The equation above is a generalization of that suggested
in~\cite{Sur1} to the case of a finite heavy hole mass.
Obviously the wave function is to be found as a plain wave propagating
along the lateral directions multiplied by a function of $x$ coordinate:
\vspace{0.05cm}
\begin{displaymath}
\Psi_\lambda ({\bf r})=\psi_\lambda(x)\,{\rm e}^{iq_y y+iq_z z}\
\vspace{0.05cm}
\end{displaymath}
with $\lambda=s,x,y,z$. Note that $\Psi_\lambda$ and $\psi_\lambda$ are
the spinors.
Without loss of generality choose $q_z=0$.
We further simplify our calculations assuming that $E_g$ and
$\Delta_{so}$ are much larger than $E$ and $\phi$.
This corresponds to the $4\times 4$ model~\cite{For1}.
Then (\ref{eq:1}) becomes:
\begin{widetext}
\vspace{-0.3cm}
\begin{equation}
\renewcommand{\arraystretch}{2.7}
\left(
\begin{array}{ll}
{m_{lh}+3 m_h\over 4m_h m_{lh}}
{d^2\over d x^2}-{q^2\over m_h}+2\alpha x-2E &
\pm i \left[{{d^2\over d x^2}-q^2\over 2m_h}
\pm {3q(m_h-m_{lh})\over 4m_h m_{lh}}
{d\over d x}+\alpha x- E \right]\\
\mp i\left[{{d^2\over d x^2}-q^2\over 2m_h}
\mp{3q(m_h-m_{lh})\over 4m_h m_{lh}}{d\over d x}+\alpha x- E
\right] & {1\over m_h}{d^2\over d x^2}
-{q^2(m_{lh}+3m_h)\over 4m_h m_{lh}}+2\alpha x-2E
\end{array}\right)\left(\begin{array}{c}\psi_x(x)\\ \psi_y(x)\end{array}
\right) = 0
\label{eq:2}
\end{equation}
\end{widetext}
where $m_h$ and $m_{lh}$ are heavy and light hole effective masses
related to Luttinger parameters as:
\begin{displaymath}
{m\over m_h}=\tilde{\gamma}_1-2\tilde{\gamma}_2\qquad
{m\over m_{lh}}={4m\gamma^2\over 3E_g}+\tilde{\gamma}_1+2\tilde{\gamma}_2.
\end{displaymath}
The upper sign in (\ref{eq:2}) corresponds to the spinor satisfying:
\vspace{-0.05cm}
\begin{displaymath}
\sigma_z\psi_x=\psi_x,\; \sigma_z\psi_y=\psi_y,\; \sigma_z\psi_z=-\psi_z,
\end{displaymath}
\vspace{-0.05cm}
while the lower does to:
\vspace{-0.05cm}
\begin{displaymath}
\sigma_z\psi_x=-\psi_x,\; \sigma_z\psi_y=-\psi_y,\; \sigma_z\psi_z=\psi_z,
\vspace{-0.05cm}
\end{displaymath}
Obviously the two sets of solutions are obtained form each other via the
time reversal transformation: $q\to -q$ and $\psi\to\psi^\star$. Therefore
we will consider only the system (\ref{eq:2}) with the upper sign.

It proves to be convenient to introduce new linear combinations:
\begin{eqnarray}
\psi_+=\psi_x+2i\psi_y, \qquad
\psi_-=-2\psi_x-i\psi_y
\end{eqnarray}
Then the equation (\ref{eq:2}) can be written in an elegant way:
\begin{equation}
\renewcommand{\arraystretch}{1.4}
\left\{
\begin{array}{l}
\delta\,(\psi_+^{\prime\prime}-q^2\psi_+)-2\lambda_l x\psi_+=qF\\
\delta\,(\psi_-^{\prime\prime}-q^2\psi_-)-2\lambda_l x\psi_-=F^\prime\\
\psi_+^\prime+2\psi_-^\prime-q (2\psi_++\psi_-) =-3F(1-\delta)^{-1}
\end{array}
\right.,
\label{eq:15}
\end{equation}
where
\vspace{-0.1cm}
\begin{displaymath}
\delta={m_l\over m_h}\ll 1,\quad {1\over m_l}=
{1\over 2} \left({3\over m_{lh}}-{1\over m_h}\right),
\quad \lambda_l={m_l\alpha\over\hbar^2},
\end{displaymath}
$F$ is an unknown function to be determined and introduced for convenience.
After rescaling coordinates and momenta:
\begin{displaymath}
x= \tilde x {\lambda_l\over 1-\delta}^{-1/3},\;\;
q=\eta {\lambda_l\over 1-\delta}^{1/3},\;\; F=\tilde F
{\lambda_l\over 1-\delta}^{1/3},
\end{displaymath}
equation (\ref{eq:15}) becomes:
\begin{equation}
\renewcommand{\arraystretch}{1.5}
\left\{
\begin{array}{l}
\delta\,(\psi_+^{\prime\prime}-\eta^2\psi_+)-2(1-\delta) \tilde x\psi_+=
\eta F\\
\delta\,(\psi_-^{\prime\prime}-\eta^2\psi_-)-2(1-\delta) \tilde x\psi_-=
F^\prime\\
\psi_+^\prime+2\psi_-^\prime-\eta (2\psi_++\psi_-) =-3F(1-\delta)^{-1}
\end{array}
\right. .
\label{eq:15'}
\end{equation}
It is convenient to perform Fourier transform and then substitute:
\begin{equation}
\renewcommand{\arraystretch}{1.6}
\left\{\begin{array}{l}
\hat{\psi}_+(k)=\tilde{\psi}_+(k)\exp{\left(i{\delta\over 2 (1-\delta)}
\left[{k^3\over 3}+\eta^2k\right]\right)}\\
\hat{\psi}_-(k)=\tilde{\psi}_-(k)\exp{\left(i{\delta\over 2(1-\delta)}\left[
{k^3\over 3}+\eta^2k\right]\right)}\\
\hat{\tilde F}(k)=G(k)\exp{\left(i{\delta\over 2 (1-\delta)}\left[
{k^3\over 3}+\eta^2k\right]\right)}.
\end{array}\right.
\label{eq:15a}
\end{equation}
Finally, we obtain a second order differential equation for the function
$G$:
\begin{equation}
{d^2 G\over dk^2}-i{k^2+\eta^2\over 3}{dG\over dk}
-\left(ik+{\eta\over 6}\right)G=0.
\label{main}
\end{equation}
The boundary condition corresponding to the incident light hole is defined
as the asymptotic at $k\to\infty\,$:
\begin{equation}
G\to A{\sqrt{\eta^2+k^2}{\rm e}^{-{i\over 2}\tan^{\!-1}{k\over \eta}}}
\exp{\left({i\over 3}\left[{k^3\over 3}+\eta^2k
\right]\right)},
\label{eq:19a}
\end{equation}
The other asymptotic at $k\to -\infty$ is:
\begin{eqnarray}
G\to && A r_{ll}{\sqrt{\eta^2+k^2}{\rm e}^{-{i\over 2}\tan^{\!-1}
{k\over \eta}}}
\exp{\left({i\over 3}\left[{k^3\over 3}+\eta^2k
\right]\right)}\nonumber\\
&&+A r_{hl}{{\rm e}^{{i\over 2}\tan^{\!-1}{k\over \eta}}
\over \left(k^2+\eta^2\right)^{3/2}},
\label{eq:19}
\end{eqnarray}
where $r_{ll}$ and $r_{hl}$ are the light-light and light-heavy hole
reflection coefficients. Below we relate them to the light and
heavy hole fluxes in real space. Similar expressions can be written
for the heavy holes to define $r_{hh}$ and $r_{lh}$:
\begin{equation}
G\to B{{\rm e}^{{i\over 2}\tan^{\!-1}{k\over \eta}}
\over \left(k^2+\eta^2\right)^{3/2}}\;\;\mbox{at}\;\; k\to\infty
\label{eq:19b}
\end{equation}
and
\begin{eqnarray}
G\to && B r_{lh}{\sqrt{\eta^2+k^2}{\rm e}^{-{i\over 2}\tan^{\!-1}
{k\over \eta}}}
\exp{\left({i\over 3}\left[{k^3\over 3}+\eta^2k
\right]\right)}\nonumber\\
&&+B r_{hh}{{\rm e}^{{i\over 2}\tan^{\!-1}{k\over \eta}}
\over \left(k^2+\eta^2\right)^{3/2}}\qquad\;\;\;
\mbox{at}\;\; k\to-\infty
\label{eq:19c}
\end{eqnarray}
Time reversal symmetry of (\ref{main}) gives simple relations:
\begin{equation}
\left\{
\begin{array}{l}
|r_{ll}|^2+r_{hl}^\star r_{lh}=1\\
r_{ll}^\star r_{hl} + r_{hl}^\star r_{hh}=0\\
\end{array}
\right. \quad\mbox{or}\quad r r^\star=I,
\label{eq:31}
\end{equation}
where $I$ is a $2\times 2$ unit matrix. Note that $r$ is not unitary
and it depends only on a single parameter $\eta$.

Let us proceed with the analysis of (\ref{main}) for the case of large
$\eta$. It appears that in this limit solving this equation is similar
to finding  a superbarrier reflection coefficient in ordinary quantum
mechanics~\cite{Pokr}. This analogy suggest shifting the variable $k$
from the real axis. For positive $\eta$ we substitute:
\vspace{-0.1cm}
\begin{displaymath}
k=i\eta+\xi
\vspace{-0.1cm}
\end{displaymath}
yielding:
\begin{equation}
{d^2 G\over d\xi^2}-i{\xi\over 3}(\xi+2i\eta){dG\over d\xi}-
\left(i\xi-{5\eta\over 6}\right)G=0.
\label{main1}
\end{equation}
The boundary condition corresponding to the incident light hole
is:
\begin{equation}
G\to A {\xi^{1/4}(2i\eta+\xi)^{3/4}}\,{\rm e}^{-{2\over 9}\eta^3
+i{\xi^3\over 9}-{\xi^2\eta\over 3}}\;\;\mbox{at}\; \xi\to\infty.
\label{eq:17}
\end{equation}
Provided $\xi\ll \eta$, (\ref{main1}) simplifies to:
\begin{equation}
{d^2G\over d\xi^2}+{2\eta \xi\over 3}{dG\over d\xi}+{5\eta\over 6}G=0.
\end{equation}
The solution to this equation, which is formally can be
expressed through Whittaker functions, is most conveniently given via
the integral representation:
\begin{equation}
G=A{\rm e}^{-{2\over 9}\eta^3} {\rm e}^{i{\pi\over 4}}
{3^{3/4}\over \sqrt{2\pi}} \int\limits_{-\infty+i0}^{\infty+i0}
t^{1/4}{\rm e}^{-{3t^2\over 4\eta}}{\rm e}^{it\xi} dt.
\label{eq:18}
\end{equation}
The particular choice of a multiplier before the integral ensures
coincidence of (\ref{eq:18}) and (\ref{eq:17}) in the
region $\sqrt{\eta^{-1}}\ll \xi\ll \eta$. This can be verified by
applying the steepest descend integration to (\ref{eq:18}). In the
region $-\eta\ll\xi\ll -\sqrt{\eta^{-1}}$ we find asymptotics
corresponding to $r_{hl}$ and $r_{ll}$. However, as noted
above the $r_{ll}$ coefficient must be set to be $1$ and not the value
obtained in our manipulation. In fact loss of accuracy for some
branches of a solution is quite common when asymptotical expressions
are extended into the complex plain. On the other
hand the $r_{hl}$ coefficient is found with exponential accuracy:
\vspace{0.05cm}
\begin{equation}
r_{hl}\approx {\rm e}^{-{2\over 9}\eta^3}{\rm e}^{-i{3\pi\over 4}}
{3^{3/4}\over 4\sqrt\pi}(2\eta)^{7/4}\Gamma(1/4),
\vspace{0.05cm}
\label{eq:18b}
\end{equation}
where $\Gamma$ denotes the $\Gamma$-function. If $\eta$ is large and
negative, one has to shift variable to $k=-i\eta+\xi$. Coming through
the same steps we can get:
\begin{equation}
r_{hl}\approx {\rm e}^{{2\over 9}\eta^3}{\rm e}^{i{\pi\over 4}}
{3^{9/4}\over 2^{7/4}\sqrt\pi}|\eta|^{1/4}\Gamma(3/4).
\vspace{0.05cm}
\label{eq:18c}
\end{equation}

In a similar fashion we can provide analysis for small values of $\eta$.
As will be clear later it is sufficient to restrict calculations
only to the case $\eta=0$. Then (\ref{main}) has two simple solutions:
\begin{eqnarray}
G_1(k)&=&k\,{\rm e}^{i{k^3\over 9}-i{\pi\over 4}},\\
G_2(k)&=&3k\,{\rm e}^{i{k^3\over 9}+i{3\pi\over 4}}
{\rm P}\int\limits_k^\infty {1\over t^2}\,{\rm e}^{-i{t^3\over 9}} dt,
\end{eqnarray}
where ${\rm P}$ stands for the principal value of the integral.
The coefficients before $G_1(k)$ and $G_2(k)$ are chosen in the way
that at $k\to\infty$ these functions coincide with the defined above
asymptotics of the light and heavy holes respectively. It is not
difficult to observe that
\begin{equation}
\renewcommand{\arraystretch}{1.5}
\begin{array}{ll}
r_{ll}=i & r_{hl}=0\\
r_{lh}=-{i\over 3^{7/6}\phantom{X\over X}\hspace{-8pt}}
\Gamma(2/3) & r_{hh}=-i.
\end{array}
\label{eq:40}
\end{equation}
This matrix obviously satisfies conditions (\ref{eq:31}).

Now let us return to real space.
From (\ref{eq:15'}) and (\ref{eq:15a}) we obtain
\begin{eqnarray}
\psi_+(\tilde{x})\!&=&\!\eta\int\limits_{-\infty}^\infty dk\,
{\rm e}^{i{\delta\over 2(1-\delta)}\left[{k^3\over 3}+\eta^2k\right]+
ik\tilde{x}}\int\limits_{k}^\infty G(t)\, dt,\phantom{xx}\\
\psi_-(\tilde{x})\!&=&\!i\int\limits_{-\infty}^\infty dk\,
{\rm e}^{i{\delta\over 2(1-\delta)}\left[{k^3\over 3}+\eta^2k\right]
+ik\tilde{x}}\int\limits_{k}^\infty t G(t)\,dt.\phantom{xx}
\label{eq:100}
\end{eqnarray}
Integrating equation (\ref{main}) itself and multiplied by $k$, one finds:
\begin{eqnarray}
\int\limits_k^\infty\! G(t)\, dt&=&{2(\eta-2ik)\over\eta^2}
\left(G(k)\,{\rm e}^{-{i\over 3}\left({k^3\over 3}
+\eta^2 k\right)}\right)^\prime_k\!\!+i{4\over\eta^2}G(k)
\label{eq:20}\phantom{XX}\\
\int\limits_k^\infty\! t G(t)\, dt&=&{2(2i\eta+k)\over\eta}
\left(G(k)\,{\rm e}^{-{i\over 3}\left({k^3\over 3}
+\eta^2 k\right)}\right)^\prime_k\!\!-{2\over\eta}G(k).\phantom{XX}
\label{eq:21}
\end{eqnarray}
In the limit $|\eta|\gg 1$, $r_{ll}$ is close to 1 and using
(\ref{eq:19a}), (\ref{eq:20}), and (\ref{eq:21}) we can find that the
light hole component of the wave function is approximately equal to:
\begin{eqnarray}
\psi_+^0(\tilde{x})\!&\approx& {i A\sqrt{3}\over 2\sqrt{\pi}}\!
\int\limits_{-\infty}^{\infty}\!\!dk
{\eta\,{\rm e}^{-{i\over 2}\tan^{\!\!-1}\!\!{k\over\eta}}
\over \sqrt{k^2+\eta^2}}\,
{\rm e}^{i{1-{\delta\over 2}\over 3-3\delta}
\left[{k^3\over3}+\eta^2 k\right]
+ik\tilde{x}}\phantom{xx}\label{eq:22}\\
\psi_-^0(\tilde{x})\!&\approx&\! -{A\sqrt{3}\over 2\sqrt{\pi}}\!\!
\int\limits_{-\infty}^{\infty}\!\!dk
{k\,{\rm e}^{-{i\over 2}\tan^{\!\!-1}\!\!{k\over\eta}}\over \sqrt{k^2+\eta^2}}
\,{\rm e}^{i{1-{\delta\over 2}\over 3-3\delta}\left[{k^3\over3}+
\eta^2 k\right]+ik\tilde{x}}\!.\phantom{xx}
\label{eq:23}
\end{eqnarray}
\vspace{1cm}
The heavy hole component is given by
\begin{eqnarray}
\psi_x^1(\tilde{x})\!&\approx&\! {iA r_{hl}\over 3\sqrt{3\pi}\eta}\!
\int\limits_{-\infty}^0\!\! dk\,{\eta\,{\rm e}^{{i\over 2}\tan^{\!\!-1}
{k\over\eta}}\over \sqrt{k^2+\eta^2}}
{\rm e}^{i{\delta\over 2-2\delta}
\left[{k^3\over 3}+\eta^2k\right]+ik\tilde{x}}\!,
\phantom{xx}\label{eq:24}\\
\psi_y^1(\tilde{x})\!&\approx&\! {i A r_{hl}\over 3\sqrt{3\pi}\eta}\!
\int\limits_{-\infty}^0\!\! dk\,{k\,{\rm e}^{{i\over 2}\tan^{\!\!-1}
{k\over\eta}}\over \sqrt{k^2+\eta^2}}
{\rm e}^{i{\delta\over 2-2\delta}
\left[{k^3\over 3}+\eta^2k\right]+ik\tilde{x}}\!.\phantom{xx}
\label{eq:25}
\end{eqnarray}

We used $x-y$ basis for the last equations since it is more suitable for
heavy holes.
From (\ref{eq:22}) and (\ref{eq:23}) we deduce semiclassical
expressions for the light hole wavefunctions valid at large negative $x$,
provided $|\eta|\gg 1$.
\begin{widetext}
\begin{equation}
\vspace{-0.05cm}
\renewcommand{\arraystretch}{2}
\left\{\begin{array}{l}
\psi^l_+(\tilde{x})\approx {3\sqrt{3}iA\over \sqrt{k_l(\tilde{x})}}
{\eta\over\sqrt{k_l(\tilde{x})^2+\eta^2}}
\sqrt{1-\delta\over 1-\delta/2}
\cos\left(\int\limits_{\tilde{x}}^{\tilde{a}_l}
k_l(\tilde{x}^\prime)d\tilde{x}^\prime+{1\over 2}\tan^{\!-1} {k_l(\tilde{x})
\over\eta}-{\pi\over 4}\right)\\
\psi^l_-(\tilde{x})\approx {3\sqrt{3}iA\over \sqrt{k_l(\tilde{x})}}
{k_l(\tilde{x})
\over\sqrt{k_l(\tilde{x})^2+\eta^2}} \sqrt{1-\delta\over 1-\delta/2}
\sin\left(\int\limits_{\tilde{x}}^{\tilde{a}_l}
k_l(\tilde{x}^\prime)d\tilde{x}^\prime
+{1\over 2}\tan^{\!-1} {k_l(\tilde{x})\over\eta}-{\pi\over 4}
\right)
\end{array}.\right.
\label{eq:50}
\end{equation}
Here $\tilde{a}_l$ is a classical turning point and
$k_l(\tilde{x})$ is a position-dependent light hole wavevector:
\begin{displaymath}
\vspace{-0.1cm}
\tilde{a}_l=-\eta^2 {1-\delta/2\over 3(1-\delta)},\;\;
k_l(\tilde{x})=\sqrt{-\eta^2-\tilde{x} {3(1-\delta)\over 1-\delta/2}}.
\end{displaymath}
\end{widetext}
Similarly we can deduce semiclassical wave functions valid for $\eta\ll 1$.
The only difference, which appears as compared to (\ref{eq:50})
is the absence of ($\tan^{\!-1}{k_l/\eta}$) in the argument of cosine.

We note that at $\delta\to 0$ expressions
(\ref{eq:24}) and (\ref{eq:25}) diverge at $x\to 0$. In fact, as was noted
in~\cite{Suris}, this phenomenon is closely related to the divergence of
the electric field in electromagnetic waves incident to the media with
linearly increasing dielectric constant~\cite{Gins}. It can be shown
that the system (\ref{eq:15}) is equivalent to the fourth order
differential equation
with a small coefficient at a higher derivative. And the equations
under consideration have precisely the form of the Maxwell equations for
the electric and magnetic fields. The parameter $\delta$ is analogous
to the coupling between plasmons and the electromagnetic field. In other
words, the effect of heavy holes on the light hole wave functions is
similar to account of spatial dependence of dielectric constant in plasma.
However, in order to get a sharp resonance, $\delta$ should be very small.
For real semiconductors its value is usually bounded between $10^{-2}$
and $10^{-1}$, which appears to be insufficient to observe the peak. On
the contrary, for plasma the corresponding quantity
($\delta_p\approx T/mc^2$) is several orders of magnitude less~\cite{Gins},
and hence the resonance can exist.

In a similar manner WKB wave functions for the heavy holes can be found.
Thus for $|\eta|\gg 1$ in a classically allowed region we get:
\begin{widetext}
\begin{equation}
\renewcommand{\arraystretch}{2}
\left\{\begin{array}{l}
\psi^h_x(\tilde{x})\approx  {iB\over \sqrt{k_h(\tilde{x})}}
{\eta\over\sqrt{\eta^2+k_h(\tilde{x})^2}}
\cos\left(\int\limits_{\tilde{x}}^{\tilde{a}_h} k_h(\tilde{x}^\prime)
d\tilde{x}^\prime -{1\over 2}\tan^{\!-1}{k_h(\tilde{x})\over \eta}
-{\pi\over 4}\right)\\
\psi^h_y(\tilde{x})\approx {B\over \sqrt{k_h(\tilde{x})}}
{k_h(\tilde{x})\over \sqrt{\eta^2+k_h(\tilde{x})^2}}
\sin\left(\int\limits_{\tilde{x}}^{\tilde{a}_h} k_h(\tilde{x}^\prime)
d\tilde{x}^\prime-{1\over 2}\tan^{\!-1}{k_h(\tilde{x})\over \eta}
-{\pi\over 4}\right).
\end{array}
\right. ,
\label{eq:8}
\end{equation}
where
\begin{displaymath}
\tilde{a}_h=-{\delta\over 2(1-\delta)}\eta^2,\quad
k_h(\tilde{x})=\sqrt{-\eta^2-{2(1-\delta)\over \delta}\tilde{x}}.
\vspace{0.1cm}
\end{displaymath}
\end{widetext}
Note that the turning point for heavy holes ($\tilde{a}_h$) is closer to
the origin then that for the light holes as should be expected.

The next step in our analysis is to relate the $r$-matrix in momentum space
defined in (\ref{eq:19}) to the $R$-matrix in real space. To do
this we have to define the basis functions for the incident and
reflected waves. Like we have observed already, it is convenient to
work in the $+,-$ representation for the light holes and $x,y$
representation for the heavy holes.
\begin{widetext}
\begin{eqnarray}
\psi_l^\rightleftarrows &=& {3\over 2\sqrt{k_l(\tilde{x})} }
\left(
\begin{array}{l}
{i\eta\over \sqrt{k_l^2(\tilde{x})+\eta^2}}\\
\mp{k_l(\tilde{x})\over \sqrt{k_l^2(\tilde{x})+\eta^2}}
\end{array}
\right)_{\!+,-}\!\!\!\! \,{\rm e}^{\mp i\int\limits_{\tilde{x}}^{\tilde{a}_l}
k_l(\tilde{x}^\prime)d\tilde{x}^\prime\mp {i\over 2}
\tan^{\!-1}{k_l(\tilde{x})\over \eta}
\pm i{\pi\over 4}}\label{eq:30}\\
\psi_h^\rightleftarrows &=&  {1\over \sqrt{2\delta k_h(\tilde{x})} }
\left(
\begin{array}{l}
{\eta\over \sqrt{k_h^2(\tilde{x})+\eta^2}}\\
\mp {k_h(\tilde{x})\over \sqrt{k_h^2(\tilde{x})+\eta^2}}
\end{array}
\right)_{\!x,y}\!\!\!\! \,{\rm e}^{\mp i\int\limits_{\tilde{x}}^{\tilde{a}_h}
k_h(\tilde{x}^\prime)d\tilde{x}^\prime\pm{i\over 2}
\tan^{\!-1}{k_h(\tilde{x})\over \eta}
\pm i{\pi\over 4}},\label{eq:30b}
\end{eqnarray}
where $\rightleftarrows$ refer to the incoming and outgoing waves.
The $x$-components of the fluxes associated with these basis
wavefunctions are equal to:
\begin{eqnarray}
&&j_{\,l}^\rightleftarrows
\!=\!{\hbar\over m_l}\Im\left[\psi_{l,x}{d \psi^\star_x\over dx}
+q\psi_{l,x}\psi^\star_{l,y} \right]\!\!
=\pm{\hbar\tilde{\lambda}^{1\over3}\over m_l}
\label{eq:26},\\
&&j_h^\rightleftarrows
\!=\!{\hbar\over m_h}\Im\left[\psi_{h,x}{d \psi^\star_{h,x}\over dx}
-q\psi_{h,x}\psi^\star_{h,y}+2\psi_{h,y}{d \psi^\star_{h,y}\over dx}
\right]\!\!=\pm {\hbar\tilde{\lambda}^{1\over 3}\over m_l}.
\label{eq:27}
\end{eqnarray}
\end{widetext}
Here we explicitly inserted  $\tilde{\lambda}_l$ and $\hbar$.
The choice of the basis functions (\ref{eq:30}) and (\ref{eq:30b}) is now
evident, since they carry out equal fluxes. Now it is not hard to relate
the $R$ and $r$ matricies:
\begin{equation}
\renewcommand{\arraystretch}{2.5}
\left( \begin{array}{cc}
R_{ll} & R_{lh}\\
R_{hl} & R_{hh}\end{array}
\right)=
\left( \begin{array}{cc}
r_{ll} & {3\sqrt{3}\eta\over 2\sqrt{1-\delta/2}}r_{lh}\\
{2\sqrt{1-\delta/2}\over 3\sqrt{3}\eta} r_{hl} & r_{hh}\end{array}
\right).
\label{eq:32}
\end{equation}
The unitarity condition for $R$ requires that
\begin{equation}
|r_{ll}|^2+{4\over 27 \eta^2}|r_{hl}|^2=1,
\end{equation}
where we neglected by $\delta$ compared to $1$. All other
relations are just a consequence of (\ref{eq:31}), in particular:
\begin{displaymath}
|R_{hl}|=|R_{lh}|,\quad |R_{hh}|=|R_{ll}|.
\end{displaymath}

Using (\ref{eq:18b}), (\ref{eq:18c}) and (\ref{eq:32}) we can find the
approximate expression for the light-heavy hole transformation coefficient
$|R_{hl}|^2$ valid for large $\eta$ .
\begin{eqnarray}
|R_{hl}|^2 &\sim&
{1\over \pi}{2\sqrt{2}\over 3\sqrt{3}}\, \eta^{3/2}\Gamma^2(1/4)
\,{\rm e}^{-{4\over 9}\eta^3}\;\;\mbox{at}\;\;
\eta\to\infty, \label{eq:28}\\
|R_{hl}|^2 &\sim&
{1\over \pi}{3\sqrt{3}\over 2\sqrt{2}}\, {1\over \eta^{3/2}}\Gamma^2(3/4)
\,{\rm e}^{{4\over 9}\eta^3}\;\;\mbox{at}\;\;
\eta\to-\infty, \label{eq:28a}
\end{eqnarray}
At small $\eta$, from (\ref{eq:40}) and (\ref{eq:32}) we obtain:
\begin{equation}
|R_{hl}|^2=|R_{lh}|^2\sim {3^{2/3}\eta^2\over 4}
\Gamma^2(2/3)\quad\mbox{at}\;\eta\to 0.
\label{eq:29}
\end{equation}
Using asymptotics (\ref{eq:28}), (\ref{eq:28a}) and
(\ref{eq:29}) as well as the
exact numerical results it is possible to find an interpolating
function which gives an excellent description of $|R_{hl}|^2$ for
all values of $\eta$ (see Figure 1):
\begin{eqnarray}
|R_{hl}|^2\!&\approx& {\eta^2{3^{2/3}\over 4}\Gamma^2(2/3)\,
{\rm e}^{\,{1.3}\eta-{4\over 9}\eta^3}
\over \sqrt{\eta}\,{9\,3^{1/6}\pi\Gamma^2(2/3)
\over 8\sqrt{2}\,\Gamma^2(1/4)}
\,{\rm e}^{\,{1.3}\eta}+1}\qquad\quad \mbox{at}\;\eta>0,\label{eq:38a}\\
|R_{hl}|^2\!&\approx& {\eta^2{3^{2/3}\over 4}\Gamma^2(2/3)\,
{\rm e}^{{4\over 9}\eta^3} \over \eta^{7/2}{3^{1/6}\pi\Gamma^2(2/3)
\over 3\sqrt{2}\, \Gamma^2(3/4)}\,
+{\rm e}^{\,1.6|\eta|^{1.5}+{4\over 9}\eta^3}}\;\mbox{at}\;\eta<0.
\phantom{XX}
\label{eq:38}
\end{eqnarray}

\section{Discussion}
From the expression (\ref{eq:38}) it follows that the light-heavy hole
transformation coefficient vanishes both in the limit
of small and large $\eta$. The maximums occur at $\eta$ close to $\pm 1$,
see Figure 1. Note that the maximum reflection coefficient at positive
$\eta$ is much larger than that at negative. An important result is that the
dependence of the reflection matrix $R$ (and transformation
coefficient $R_{hl}$ in particular) on the
dimensionless longitudinal momentum $\eta$ is not sensitive
to the ratio of the light and heavy hole masses. Also the potential
slope completely scales out and the crossover from small to large $q$
occurs even in the case of a very slow varying potential.

\begin{figure}
\label{fig1}
\includegraphics[angle=90, width=8.6cm, height=6.5cm]{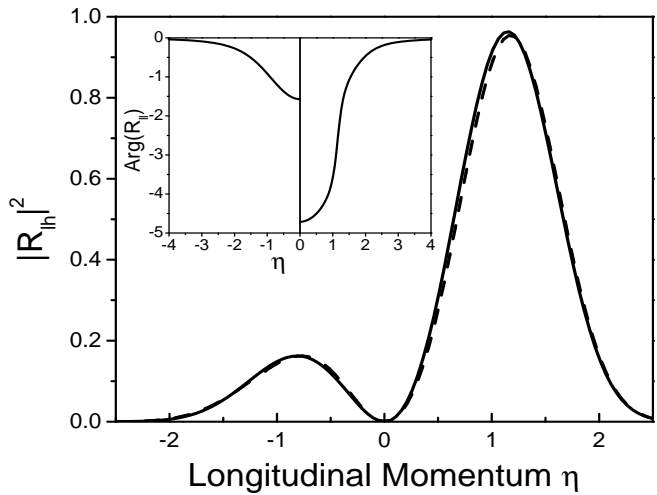}
\caption{Dependence of the light-heavy hole transformation coefficient
$|R_{hl}|^2$ on $\eta$, obtained from numerical solution of (\ref{main})
- solid line, and the approximate expressions (\ref{eq:38a}),
(\ref{eq:38}) - dotted line.
The insert shows the relative phase of the light hole reflection
coefficient $R_{ll}$.}
\end{figure}

In the insert of Fig. 1 we plot the phase of the light hole reflection
coefficient. It is equal to the conventional value for large $\eta$.
The discontinuity of the phase at $\eta=0$ is due to inclusion of
$\pm\tan^{\!-1} (k/\eta)$ into the basis functions (\ref{eq:30}) and
(\ref{eq:30b}). Similar curve but with the opposite sign is valid for
the heavy holes.

Now let us compare transformation coefficient studied in this paper with
that obtained for a scattering on an abrupt potential. There is no unique
and justified approach to derive the boundary conditions for envelope
functions~\cite{Kisin}. However, for the infinite barrier the most natural
and used requirement is the vanishing of wave function components at the
interface (see for example refs~\cite{Sercel, Efros, Chuang}).
Using these boundary conditions and basis functions (\ref{eq:30})
and (\ref{eq:30b}) with $k_l$ and $k_h$ being constants independent
of $x$, it is not hard to derive reflection matrix for the scattering
from an infinite abrupt potential. In particular one finds:
\begin{eqnarray}
|R_{hl}^a|^2=|R_{lh}^a|^2
&=&{12 k_lk_h q^2\over 4(k_lk_h+q^2)^2+q^2(k_h-k_l)^2},\\
|R_{hh}^a|^2=|R_{ll}^a|^2&=&{4(k_lk_h-q^2)^2+q^2(k_h+k_l)^2\over
4(k_lk_h+q^2)^2+q^2(k_h-k_l)^2}.
\label{eq:50a}
\end{eqnarray}
Introducing the incidence angle for the light holes:
\begin{displaymath}
\sin\phi={q\over k_l^2+q^2}
\end{displaymath}
and using $\delta\ll 1$ we get:
\begin{equation}
|R_{hl}^a|^2= {6\sqrt{6\,\delta}\sin^2\phi\cos\phi\over
(\sin^2\phi +4\cos^2\phi)}+O(\delta).
\label{eq:51}
\end{equation}
In a more general case of nonspherical band ($\gamma_2\neq \gamma_3$)
light and heavy hole transformation coefficients were numerically
investigated in~\cite{Chuang}.
Expression (\ref{eq:51}), contrary to (\ref{eq:38}), has explicit
dependence on $\delta$. It vanishes in both limits of normal and
lateral incident light hole. However, as we mentioned above, use of
vanishing at interface boundary conditions is controversial and needs
further justification. Note that the reflection matrix from a linear
smooth potential is found without making any specific assumptions
about semiconductors. So the result of the present analysis
is quite general and doesn't depend on microscopical details.

\begin{figure}
\label{fig2}
\includegraphics[angle=90, width=4.25cm, height=3.25cm]{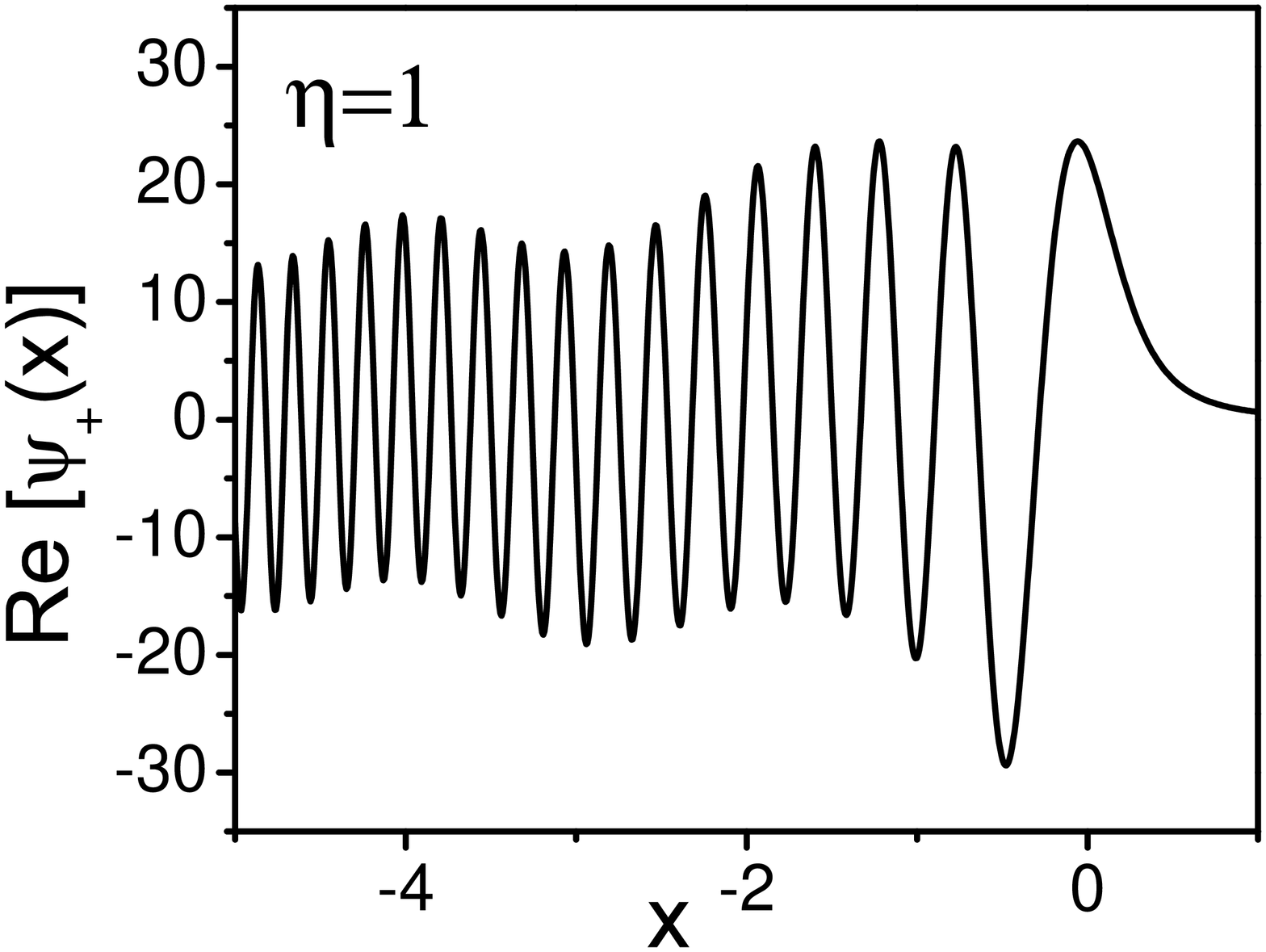}
\includegraphics[angle=90, width=4.25cm, height=3.25cm]{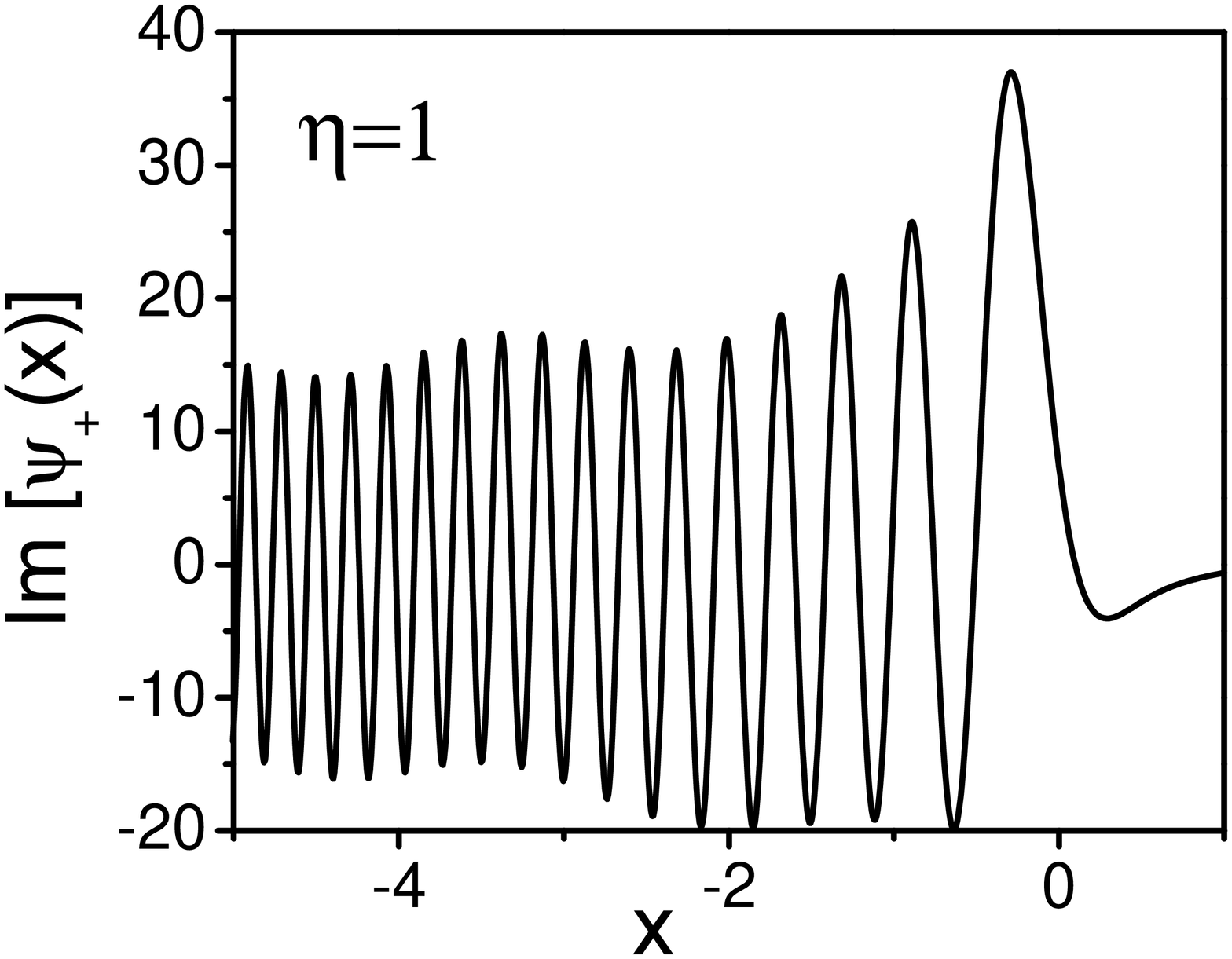}\\
\includegraphics[angle=90, width=4.25cm, height=3.25cm]{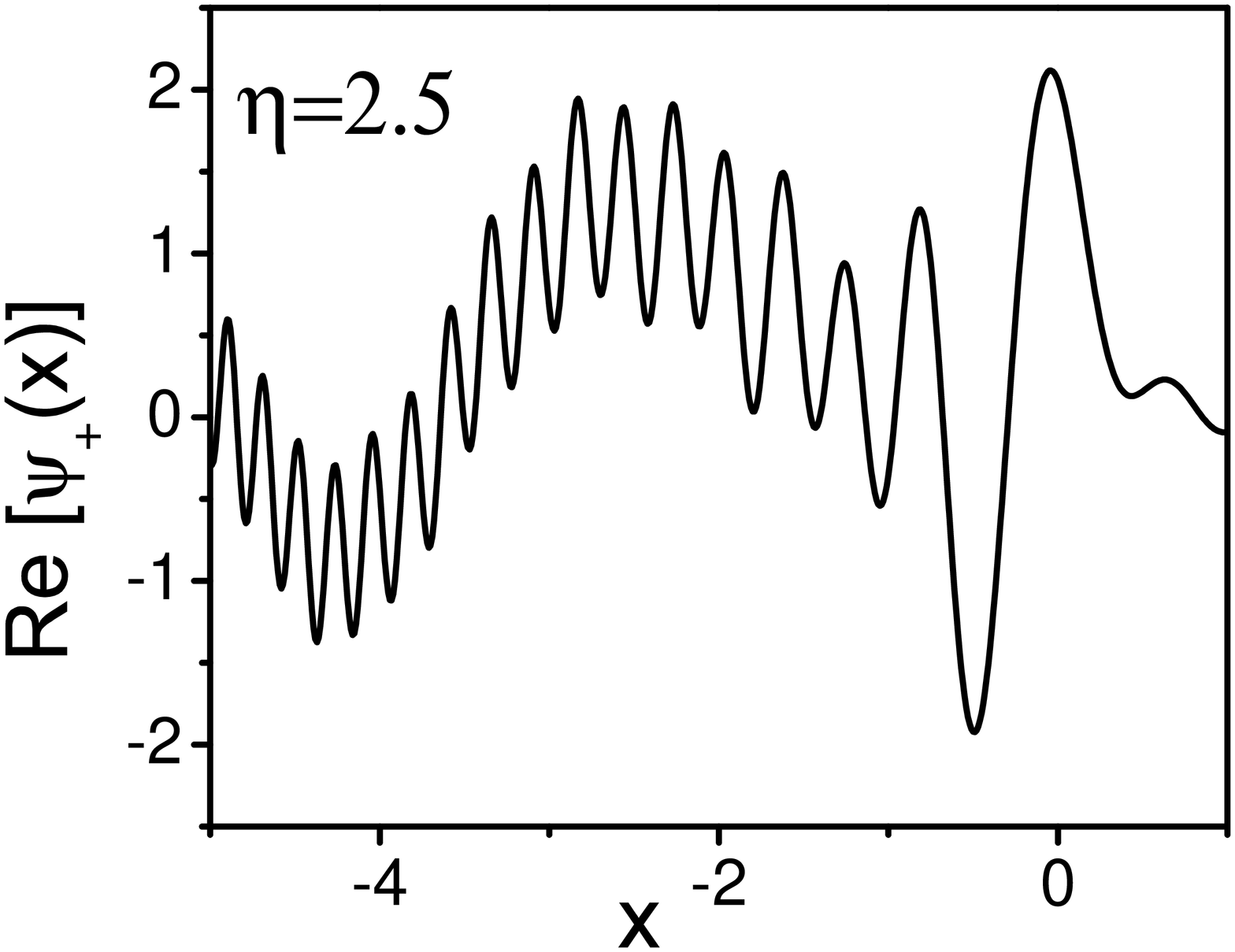}
\includegraphics[angle=90, width=4.25cm, height=3.25cm]{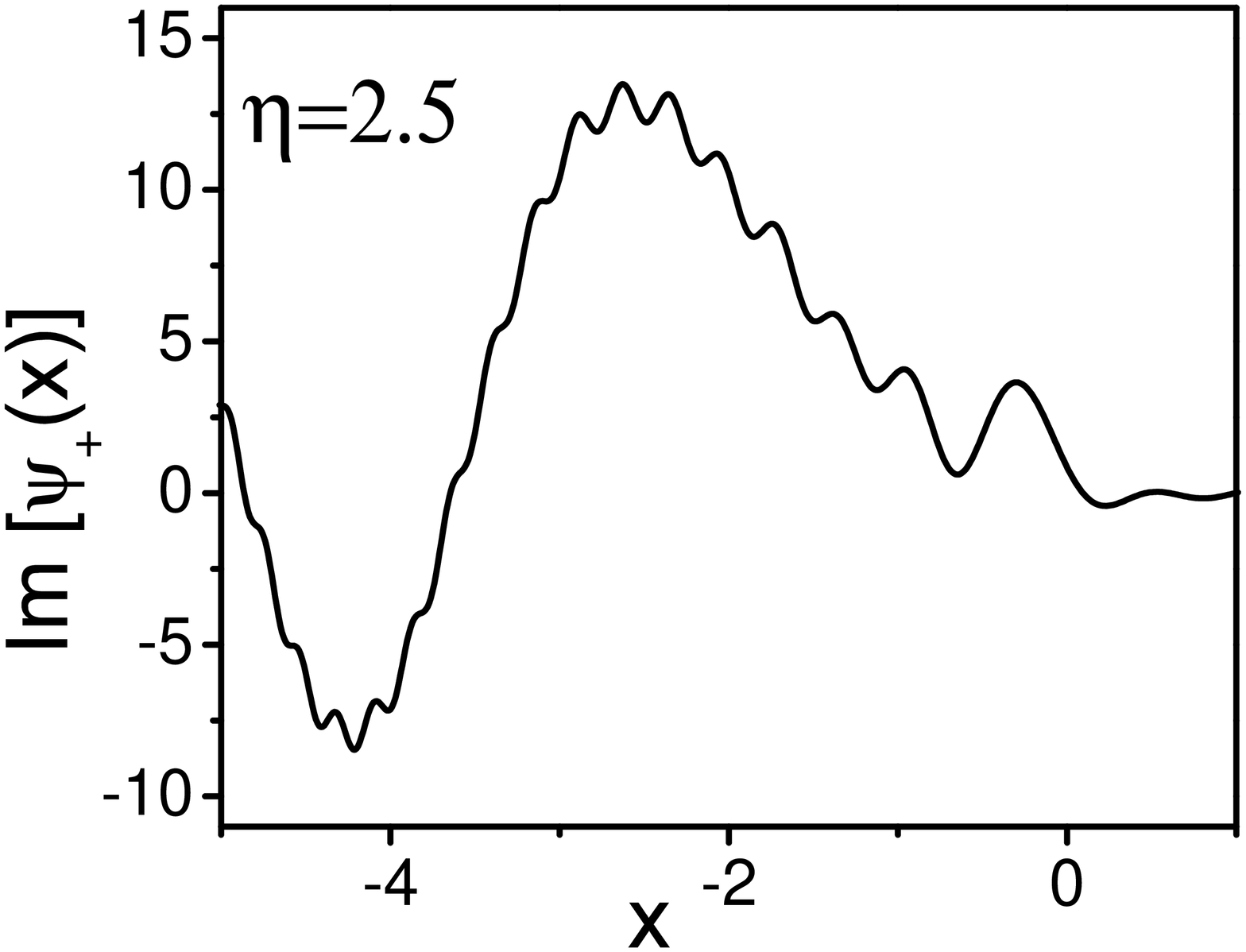}
\caption{Real and imaginary part of $\psi_+$ as a function of
$x$-coordinate for the case of the incident light hole. The top graphs
refer to the large mixing between the hole branches ($\eta=1$) and
the bottom graphs do to the small mixing ($\eta=2.5$).}
\end{figure}

Figure 2 shows the wave functions for the incident
light hole for two different values of $\eta$ corresponding to
large and small mixing between the two hole branches. High frequency
oscillations correspond to the heavy hole component, which is obviously
essential for $\eta=1$ and relatively small for $\eta=2.5$. Note
that as was mentioned earlier at $\delta\to 0$ there appears a
sharp resonance in the wave function near $x=0$. However it is hardly
seen in the graphs.
The reason is that the height of the peak, as it follows from (\ref{eq:24},
\ref{eq:25}),  is of the order $\delta^{-1/3} \exp(-2/9\, \eta^3)$ and the
peak itself exists only when $|\eta|\gg 1$. Real values of $\delta$
for the semiconductors are not less then $10^{-2}$, which is insufficient
for observing the resonance.

In conclusion, the light-heavy hole transformation coefficient for
reflection from a linear potential is found to be a function of a
dimensionless parameter $\eta$, which is proportional to the longitudinal
momentum of the light holes, and to be independent of the ratio of the
light and heavy hole masses if this ratio is small. This function vanishes
in both limits of large and small $\eta$, however the phases of light-light
and heavy-heavy hole reflection coefficients are different in these limits.
An approximate analytical expression for the transformation coefficient is
found (see (\ref{eq:38})). Account of the finite heavy hole mass is shown
to be responsible for disappearance of an underbarrier resonance in the
light hole wave function predicted in~\cite{Suris}.

After the paper was written authors became aware of the work~\cite{Perel},
where some of the results, in particular asymptotical expressions
(\ref{eq:28}) and (\ref{eq:28a}), were already obtained.

This work was partially supported by the Russian Foundation for
Basic Research. One of the authors is grateful to
I. Yu. Solov'ev for useful discussions.

\end{document}